\documentclass[sigconf, nonacm, 11pt]{acmart}

\usepackage{graphicx}
\usepackage{hyperref}
\usepackage{cleveref}
\usepackage{tabulary}
\usepackage{soul}
\usepackage{subcaption}
\usepackage{algorithm}
\usepackage{algorithmic}
\usepackage{listings}
\usepackage{xcolor}
\usepackage{multirow}

\AtBeginDocument{%
  \providecommand\BibTeX{{%
    \normalfont B\kern-0.5em{\scshape i\kern-0.25em b}\kern-0.8em\TeX}}}

\usepackage{ifthen}
\usepackage[normalem]{ulem} 

\newboolean{showedits}
\setboolean{showedits}{true} 
\ifthenelse{\boolean{showedits}}
{
	\newcommand{\del}[1]{\textcolor{red}{\sout{#1}}} 
}{
	\newcommand{\del}[1]{} 
	
}

\newboolean{showcomments}
\setboolean{showcomments}{true}
\newcommand{\id}[1]{$-$Id: scgPaper.tex 32478 2010-04-29 09:11:32Z oscar $-$}

\ifthenelse{\boolean{showcomments}}
{\newcommand{\nbc}[3]{
 {\colorbox{#3}{\bfseries\sffamily\scriptsize\textcolor{white}{#1}}}
 {\textcolor{#3}{\sf\small$\blacktriangleright${#2}$\blacktriangleleft$}}}
 }
{\newcommand{\nbc}[3]{}
 \renewcommand{\del}[1]{} 
 }

\definecolor{ibcolor}{rgb}{0.4,0.6,0.2}

\usepackage{wasysym}

\definecolor{aycolor}{rgb}{0.2,0.4,0.2}

\definecolor{ascolor}{rgb}{0.96,0.02,0.83}				

\definecolor{hccolor}{rgb}{0.21,0.54,0.84}

\definecolor{ideacolor}{rgb}{1.0,0,0.5}

\definecolor{mccolor}{rgb}{0.2,0.2,0.6}

\definecolor{abstractcolor}{rgb}{0.0,0.5,1.0}

\definecolor{introcolor}{rgb}{0.0,1.0,0.5}

\definecolor{papercolor}{rgb}{1.0,1.0,0.0}

\definecolor{multicolor}{rgb}{1.0,0,0}

\definecolor{todocolor}{rgb}{0.9,0.1,0.1}

\definecolor{qcolor}{rgb}{0.2,0.0,0.9}

\begin{document}

\title[Systems for Memory Disaggregation: Challenges \& Opportunities ]
{Systems for Memory Disaggregation: Challenges \& Opportunities} %
\author{Anil Yelam}
\email{ayelam@ucsd.edu}

\settopmatter{printfolios=true}
\sloppy

\begin{abstract}
Memory disaggregation addresses memory imbalance in a 
cluster by decoupling CPU and memory allocations of 
applications while also increasing the effective memory 
capacity for (memory-intensive) applications beyond the 
local memory limit imposed by traditional fixed-capacity 
servers. As the network speeds in the tightly-knit 
environments like modern datacenters inch closer to the 
DRAM speeds, there has been a recent proliferation of 
work in this space ranging from software solutions that 
pool memory of traditional servers for the shared use of 
the cluster to systems targeting the memory disaggregation
in the hardware. 
In this report, we look at some of these recent memory 
disaggregation systems and study the important factors 
that guide their design, such as the interface through 
which the memory is exposed to the application, their  
runtime design and relevant optimizations to 
retain the near-native application performance, 
various approaches they employ in managing cluster memory 
to maximize utilization, etc. and we analyze the associated 
trade-offs. We conclude with a discussion 
on some open questions and potential future directions  
that can render disaggregation more amenable for adoption.

\end{abstract}
\maketitle

\section{INTRODUCTION}
\label{sec:intro}

With the tremendous growth of computing in the past two 
decades, applications have become both data-intensive 
and latency-sensitive, which gave rise to in-memory 
computing in lieu of going to the disk. 
This led to memory-intensive 
applications whose memory needs on a server outweigh the 
processor needs, introducing a skew in resource usage.
However, traditional servers come with fixed processor 
and memory resources that does not allow dynamically
resizing memory. This was generally solved by swapping 
to disk but disk speeds were really slow compared to memory 
affecting performance. 
At the same time, the diversification of computing usecases 
introduced a high heterogeneity of applications (e.g., cloud 
computing) with varying memory needs in proportion to 
the CPU, leaving some of the traditional servers in a data center  
with underutilized memory and others with not enough; the result
being inefficient memory utilization in the cluster and hence,
increased cost of ownership. Decoupling memory would allow 
applications to be more elastic in their memory usage and 
improve the memory utilization of the cluster at the same time.
Memory disaggregation involves such (logical or physical) 
decoupling of memory resources in a cluster from other 
(processor) resources. 

One way to alleviate memory pressure is to scale out and 
build a distributed application that runs on multiple nodes and 
adjust itself to the memory restrictions on the individual 
nodes. Indeed, there are many platforms that provide distributed 
memory management for such applications like distributed key-value 
stores~\cite{Ousterhout2010,Lim2012,Novakovic2016,Kalia2015},
distributed shared memory (DSM) 
systems~\cite{treadmarks,dsm1,farm,gam}, etc. These systems 
provide a globally accessible interface for all servers where  
the focus is on providing fine-grained memory sharing and 
a reasonable consistency model for a distributed application. 
An alternative is to extend the private memory space 
of (single-node) applications a la 
remote swapping systems~\cite{gms,cashmere} that transparently
swap application pages to remote memory, without any notion 
of sharing across servers (i.e., their memory consistency model 
stops with cache coherence protocols on a 
single server). In this report, we focus 
on the latter kind where the stress is more on perfomant 
remote access mechanisms and efficient memory management for 
individual applications and less on memory sharing and 
consistency across servers.

As the networks become faster and technologies such as 
RDMA~\cite{farm,rocev2} arrive to commodity clusters, 
the remote access latencies are inching closer to native 
DRAM latencies (which, on the contrary, are nearing 
saturation)~\cite{Aguilera2017}, making remote memory more 
accessible for applications performance-wise~\cite{netdisagg}.
Consequently, there has been a renewed interest in 
the last half-a-decade in building remote memory 
systems~\cite{infiniswap,zswap,leap,fastswap,
legoos,kona,aifm,semeru,remregions,literdma}.
Traditional way of memory disaggregation is to pool/track  
unused memory across the cluster in software 
and use it to complement memory on the memory-hungry servers.
This is still popular, with the work on remote swapping systems 
continuing to this day~\cite{infiniswap,fastswap,zswap,leap}.
The other, more recent approach is to disaggregate the memory 
in hardware and where all the memory is decoupled from 
compute and is made available to the compute nodes via 
the network~\cite{legoos,bladedisagg1,sonuma}. 
We use the terms \textit{remote} or \textit{disaggregated} 
memory synonymously to refer to all the memory available for 
shared usage of the cluster 
whether it is pooled in software or hardware. 

In both cases, building a system 
that exposes and manages such disaggregated memory face  
similar design challenges. First, the system should decide on
the right interface to expose this memory; for example,
to either be transparent and avoid any application changes, 
or to be more expressive and provide richer functionality and 
exploit application information/semantics for performance. 
Moreover, remote access 
latencies are still an order-of-magnitude worse than local, 
so the system should implement performance optimizations 
like caching or at the least, enable applications to 
implement such optimizations to hide the impact of remote 
latencies. While providing reasonable programming model and 
minimal performance degradation for a wide range of 
applications, it should also work towards efficiently 
managing the cluster memory behind the scenes and 
maintain good memory utilization. 

In this report, we explore in 
detail, the above design 
challenges of building a system for disaggregated memory, 
in addition to some more that is expected of a holistic 
system e.g., fault tolerance, reliability, security and 
isolation, etc. through the lens of recent disaggregated/far 
memory systems. Through this analysis, we hope to 
highlight the trade-offs involved with various design 
considerations. Finally, we conclude with a discussion on 
some remaining challenges and future opportunities that 
can help make disaggregation more amenable for adoption. 
\section{Memory Disaggregation}

\begin{figure*}[h!]
    \centering
    \includegraphics[width=.9\linewidth]{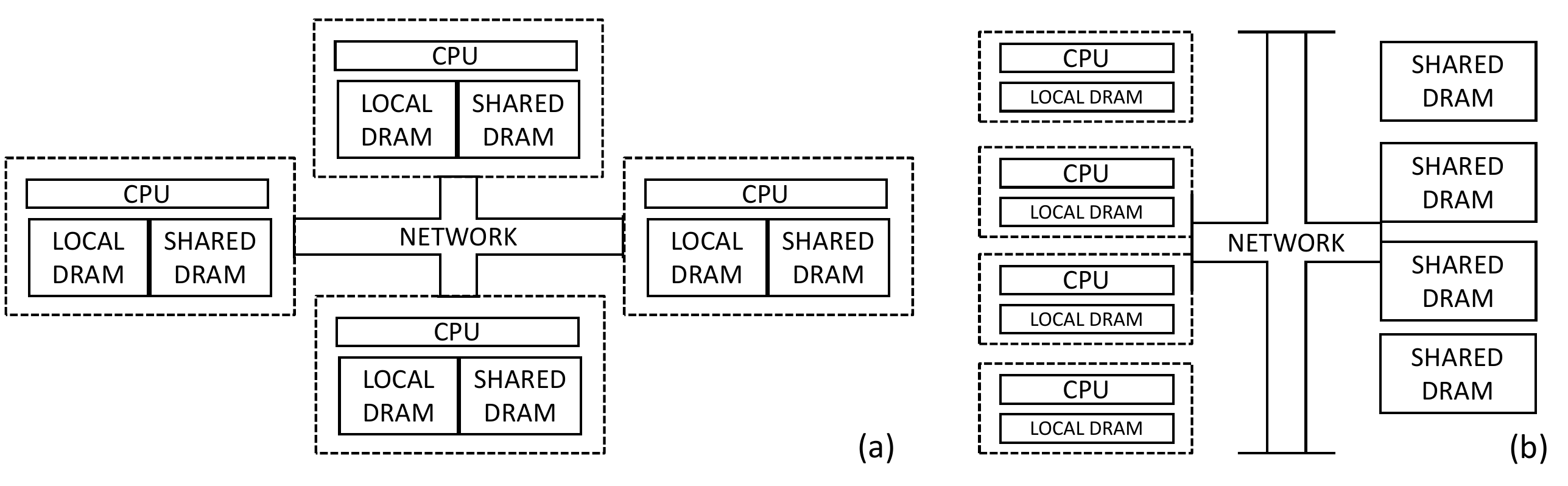}
    \caption{Shows (a) software-disaggregated 
    architecture where disaggregated memory is pooled from 
    traditional servers as opposed to the (b) hardware-disaggregated
    design where most memory is decoupled in hardware.}
    \label{fig:architecture}
\end{figure*}

Memory disaggregation aims to decouple the available compute 
and memory resources in the cluster and allow for independent 
allocations of these resources regardless of where a job 
is placed. This means, the OS/runtime 
that's running the job should provide a platform to 
expose/give access to potentially all the memory 
available in the cluster. Ideally, it should hide the 
complexity of setting up and accessing remote memory 
(e.g., RDMA connection setup) and 
expose an easy-to-use interface for working with remote memory.
At the same time, it should trade-off the properties 
of the interface with decent performance guarantees 
and other requirements from the system like resource 
sharing and isolation across applications. At a high 
level, the platform is a distributed system consisting 
of a client-side (compute-side) components (a runtime 
that exposes the memory interface and acts as an agent 
on each compute node), the server-side (memory) components 
(to manage memory on each memory server) and an 
interconnect over which these components interact to 
provide an abstraction for shared cluster memory.
It may optionally include other cluster resources like
centralized managed for global memory/metadata management
and failure handling.

\vspace{3pt}
\noindent \uline{Target Architecture.}
Proposed solutions for memory disaggregation target two 
different kind of cluster/memory architectures based on
existing technologies or technologies that are expected
to be available in the near future (shown in 
Figure ~\ref{fig:architecture}).

\vspace{3pt}
\noindent \textbf{1. Software-disaggregated.}
Some systems~\cite{gms,cashmere,infiniswap,remregions,
leap,zswap} target the traditional homogeneous 
datacenters with monolithic servers as the basic 
deployment unit, connected to each other by low-latency 
network interconnects like Infiniband or RoCE. Each 
unit hosts both compute and memory resources and the 
software provides an interface to remote memory 
on other nodes. Local memory is prioritized for 
local jobs and unutilized memory on all the nodes 
can be pooled and presented to the cluster as 
remote/disaggregated memory, which could be static 
or vary in capacity over time. 

\vspace{3pt}
\noindent \textbf{2. Hardware-disaggregated.}
Other systems~\cite{kona,aifm,fastswap,semeru,legoos} 
target a hardware disaggregated architecture where (most 
of the) memory nodes are detached from the compute nodes 
and made available through the network. The memory node 
can be a traditional monolithic server with limited 
compute and stuffed with DRAM~\cite{fastswap} or 
each DRAM unit itself directly-attached to a memory 
controller and network interface~\cite{legoos}. 
Even in a purely disaggregated setup, however, 
it is generally assumed that each compute node has 
a small amount of local memory and vice 
versa.~\cite{legoos,kona}

In both architectures, compute servers use the local memory 
to run the OS and other runtime essentials for exposing remote 
memory, and only use remote memory for the applications. 
There are many reasons for this choice. 
First, without local DRAM, all the memory 
accesses would be remote and the memory controller should 
possess the knowledge and capability to fetch remote memory 
directly without any help from software; such complex ``control 
path" knowledge would need either a ``smart" memory controller 
(e.g., RMC in soNUMA~\cite{sonuma}) or some other smart 
hardware (e.g., ccFPGA in Kona~\cite{kona}) next to it. 
Even these solutions do not put OS on remote memory and 
and maintain local DRAM to exploit cache locality as remote 
accesses are still worse than local.

{\renewcommand{\arraystretch}{1.2}
\begin{table*}[!t]
    \centering
    \begin{tabular}{l|ccccc} 
        \textbf{\shortstack[l]{Interface Type \\ (Implementation)}}
            & \textbf{System}
            & \textbf{Transparent}
            & \textbf{General}
            & \textbf{\shortstack[c]{What can be\\Remote}}
            & \textbf{\shortstack[c]{Sharing \\ support}}
        \\ \hline
        \multirow{4}{*}{\shortstack[l]{Virtual Memory \\ (Traditional Paging)}}
            & Infiniswap~\cite{infiniswap} (2017)
            & Yes
            & Yes
            & All
            & No
        \\ \cline{2-6}
            & zSwap~\cite{zswap} (2019)
            & Yes
            & Yes
            & All
            & No
        \\ \cline{2-6}
            & Leap~\cite{leap} (2019)
            & Yes
            & Yes
            & All
            & No
        \\ \cline{2-6}
            & Fastswap~\cite{fastswap} (2020)
            & Yes
            & Yes
            & All
            & No
        \\ \hline
        \multirow{2}{*}{\shortstack[l]{Virtual Memory \\ (New Hardware)}}
            & LegoOS~\cite{legoos} (2018)
            & Yes
            & Yes
            & All
            & No
        \\ \cline{2-6}
            & Kona~\cite{kona} (2021)
            & Yes
            & Yes
            & Heap
            & No
        \\ \hline
        \multirow{2}{*}{\shortstack[l]{Language-based \\ (User space)}}
            & AIFM~\cite{aifm} (2020)
            & Yes
            & No
            & Portion
            & No
        \\ \cline{2-6}
            & Semeru~\cite{semeru} (2020)
            & Yes
            & No
            & (Java) Heap
            & No
        \\ \hline
        \multirow{2}{*}{\shortstack[l]{Custom API \\ (Kernel-based)}}
            & Remote Regions~\cite{remregions} (2020)
            & No
            & Yes
            & Portion
            & Basic
        \\ \cline{2-6}
            & LITE MRs~\cite{literdma} (2017)
            & No
            & Yes
            & Portion
            & Basic
        \\ \hline
    \end{tabular}
    \vskip .5em
    \caption{Various interfaces for remote memory adopted in 
    some recent systems}
    \label{tab:interfaces}
  \end{table*}

\section{Design Factors}
\label{sec:design}
We look at various considerations that guide the design of a 
disaggregated system and for each of them, we discuss why it 
matters, how previous systems have (not) treated it, are there 
going to be trade-offs with others, etc. 

\subsection{Programming Model}
A key question for a disaggregated system is how it 
chooses to expose the memory (both local and remote) of the 
cluster to the applications running on it. Below, we talk about 
various aspects of the interface while referring to 
various interfaces proposed in previous systems 
(also listed in Table \ref{tab:interfaces}).

\subsubsection{Transparency}
Does the app (need to) know whether an access is local or 
remote? Does it require a rewrite of applications or can 
existing applications port to it with minimal or no effort?

\vspace{3pt}
\noindent \uline{Virtual memory-based transparency.}
In a traditional (x86) server, access to local memory is 
provided through the virtual memory abstraction using 
the load/store-style instructions on virtual addresses 
whose translation is handled by hardware (TLB, MMU) and 
managed by the OS. The same virtual memory abstraction can 
be extended to remote memory as well. The abstraction 
allows the actual pages to be backed my any device 
(disk, remote memory, files, etc.) from where they can 
be fetched when accessed by hooking into the page fault 
handler. The main benefit is complete backwards-compatibility 
and language-agnosticism where all the applications 
targeting x86 hardware can utilize  
remote memory without a single line of code change. The flip side 
is that this interface is rigid (it cannot be extended like a
regular API) and kernel-based, and so doesn't allow any 
app-specific information to percolate into the runtime limiting
its performance optimizations, as we will see in 
section \ref{sec:performance};

Due to its complete transparency, it has been the 
go-to interface for all the \textit{remote paging} systems 
from the old~\cite{gms,cashmere} that continue to this
day~\cite{infiniswap,fastswap,zswap,leap}.
These systems target traditional servers 
(software-disaggregation) and provide remote memory as a swap 
space by hooking into the virtual memory manager in the kernel.
In the hardware-disaggregated setting, LegoOS~\cite{legoos}, a 
operating system designed for this setting, provides a similar 
interface for disaggregated memory. LegoOS models local DRAM as
next level \textit{virtually-indexed} cache and moves address 
translation hardware (TLB, paging hardware, etc.) to the memory
nodes. Similar to page-faults, cache misses are handled in 
software when the remote pages are fetched into local DRAM.
This interface also allows the whole application memory 
(code, stack and heap) to be in remote memory as all of it 
can be transparently paged out.

Kona~\cite{kona} is a recent work that proposes a hardware-based 
implementation for this interface that avoids software overhead in
handling page faults/cache misses. It proposes hardware primitives 
that assist with trap remote memory accesses at the memory controller 
hardware and fetch them from the remote memory. In an example 
implementation, Kona uses cache-coherent FPGA that is connected to 
CPU using interconnects like CXL\cite{ccix}. This 
interconnect provides the FPGA with visibility to all the memory 
accesses through cache coherence protocol and routes all remote 
accesses through this hardware, which implements the runtime. 
This approach however requires special hardware.

\vspace{3pt}
\noindent \uline{Language-based Transparency.}
Without special hardware support like Kona~\cite{kona}, 
only a kernel (paging) based implementation can provide 
the completely transparent memory interface that can be 
exploited by all the traditional applications
without application changes. However, kernel-based 
implementations suffer from incomplete information on 
memory access patterns and data amplification overheads 
because the data tracking and movement granularity 
(fetching remote data) is restricted by the virtual memory 
system i.e., the kernel must fetch the entire page (4 KB) 
just to access even a small object. And the simplicity  
of the virtual memory interface 
(malloc and load/store) also means that it makes it 
harder to provide application-specific semantics or 
hints to the runtime for optimized implementation. 
For example, data structures designed with far memory 
awareness might perform a lot better than traditional ones 
running on a memory-transparent interface~\cite{Aguilera2019}. 

To imbibe application semantics into the runtime, systems 
like AIFM (Application-Integrated Far Memory)~\cite{aifm} 
and Semeru~\cite{semeru} opt for implementing their runtime in 
Userspace and bypass the kernel. Since native addresses can 
only point to local memory, another level of indirection 
is needed to address memory in order to hide remote memory
and retain some transparency. AIFM provides such indirection 
using C++ smart pointers while Semeru 
uses Java virtual addresses. This indirection also lets 
the runtime to track accesses at a much finer object 
granularity to perform more precise hotness tracking 
(leading to better cache eviction policies) and avoid 
data amplification. The indirection may however add some 
performance overhead in critical path compared to native 
memory accesses. Also, user space runtimes cannot place entire 
memory remotely (e.g., local process segments like stack and 
code has to be in local memory) which may offset the 
decoupling benefits of disaggregated memory.

Programming languages provide inbuilt implementations of 
common data structures like lists and hash tables either as 
language primitives or as a part of standard libraries. 
One way to take achieve app-runtime codesign while preserving 
transparency (i.e., no or minimal changes) for applications 
is to modify just these implementations under the hood to be 
far memory-aware. AIFM, for example, provides remoteable 
alternatives to standard data structures that provide access 
hints to the prefetcher or offload data-intensive operations 
like copy, aggregation, etc. to the memory server. 
Similarly, language runtimes can optimize their implementation 
around remote memory. For example, remote access latency can 
be hidden by maintaining lightweight threads and running other
threads while some thread waits for remote data~\cite{aifm}.  
Similarly, garbage collectors in managed runtimes like Java
can be optimized to target remote memory by offloading the 
data-intensive parts like object traversal to the memory 
servers~\cite{semeru}.

\vspace{3pt}
\noindent \uline{No Transparency.}
Some interfaces provide remote memory through a custom API 
(usually implemented as a set of library or system calls) and  
are not limited by the requirement to abstract away remote 
memory and to be backwards-compatible. Without such a 
limitation, the API can choose to be very expressive.
These APIs generally provide methods/calls for allocating and 
accessing remote memory, and in some cases, synchronization or 
transactional primitives for sharing memory across machines. 
Performance optimizations like caching are generally left to 
the applications. With non-transparent interfaces, however,
the choice of what goes in local vs remote memory is left to 
the application. Since local memory is inflexible, leaving 
this choice to applications may hurt the ability of the 
runtime to efficiently perform memory decoupling and other 
goals of memory disaggregation.

Examples for such interfaces include Remote 
Regions~\cite{remregions} and LITE Memory regions~\cite{literdma} 
which expose an expressive
kernel-based remote memory API in an effort to provide a 
higher-level abstraction for RDMA.
These systems provide a namespace for (contiguous) memory 
segments (of arbitrary sizes) exposed by machines across the 
cluster and allow client apps to bind to these segments, 
and perform read/write through the ioctl stubs. The expressive 
interface gives them the flexibility to expose various 
additional operations to applications like the RPC support in 
LITE and caching/prefetching hints in Remote Regions. 
These interfaces however does not take all the complexity of 
careful memory management and performance optimizations but
onloads some to the application. Other complex systems can 
certainly use these interfaces as a backend 
(e.g., LegoOS uses LITE as the interconnect) for ease of 
implementation. Other (user space) examples include the 
read/write API with transactional semantics provided by
remote memory by recent distributed computing platforms like
FARM~\cite{farm} and GAM~\cite{gam}.


\subsubsection{Generality} 
Is the exposed interface (ABI/API) 
general enough for adoption across wide range of 
platforms/applications, or does it favor a particular app 
above or particular runtime below, potentially falling out 
of favor with new kinds of apps/hardware? For example,
interfaces provided through the kernel-based runtimes 
(either transparent ones based on virtual memory or 
non-transparent ones exposed through ioctl stubs) are more 
general in the sense that they're available to  
all the applications regardless of the language they are 
written in. On the other hand, in addition to being 
language-specific, user space runtimes~\cite{aifm,semeru} 
only support remote memory 
access/management for a single application and hence cannot 
co-ordinate sharing (and isolation) of available remote 
memory among multiple applications. Such sharing requires 
mediation of the kernel, at least in the control path.

\subsubsection{Ease of programming.} 
How easy is it for applications to program with this interface? 
When working with remote memory, this depends on the amount of the 
complexity of implementation that the interface and the runtime 
hide away from the application. Naturally, transparent interfaces
are usually the easiest to program with. Of the non-transparent 
ones, the complexity may vary depending on how high- or 
low-level the abstractions they provide are. For example, 
as pointed out in~\cite{Aguilera2017}, 
adding two variables in disaggregated memory
would be a very simple operation with virtual memory interface 
(*c = *a + *b). With non-transparent but still high-level 
abstractions like LITE~\cite{literdma}, we need to first open
the remote memory as LITE region (\textit{LT\_map()}), 
read the variables using \textit{LT\_read()} and 
write the result back using \textit{LT\_write()}.
Doing the same with using RDMA would be even more complex with 
setting up queue pairs and memory regions, and reading/writing 
data by posting work requests. A common but imperfect metric 
to compare ease of programming is the number of lines of code (LOC). 
For example, both LITE~\cite{literdma} and Remote Regions~\cite{remregions} show 
two orders of magnitude reduction in LOC compared to pure 
RDMA-based implementation for various applications accessing 
remote memory.

\subsection{Application Performance}
\label{sec:performance}
When running on a disaggregated system, we ideally expect 
no or minimal degradation in application performance 
when compared to the native performance. Depending 
on the application, one can look at its job completion time, 
throughput or tail latencies as a proxy for performance. 
(Although, metrics like total cost of ownership (TCO) are 
more holistic and account for things like memory 
utilization across the cluster, special hardware costs, etc.
but they are harder to measure?). In this section, we 
look at various factors that affect the memory performance,
through these metrics.

\subsubsection{Caching}
Remote accesses across the network today are still on 
the order of magnitude slower (or worse, depending on the 
networking stack and interconnect), so application 
performance would be terrible if all accesses were 
to be remote~\cite{netdisagg}. Fortunately, most 
applications exhibit spatial and temporal locality 
in accesses which can be exploited by caching 
remote data, and as such the quality of caching can 
greatly affect the performance. Depending on the 
interface exposed, the runtime may choose to do 
caching or leave it to the application itself. 
Custom interfaces can benefit from application-specific 
caching hints. For example, AIFM~\cite{aifm} allows 
applications to exclude specified data from cache 
and avoid cache pollution. 

\vspace{3pt}
\noindent \uline{Cache Block Size.} Fetching remote data in 
bigger chunks will help exploit spatial locality 
however it also runs the risk of bringing in redundant data 
and polluting the cache, so it needs to be properly 
balanced for the best cache hit ratio. Traditional
virtual memory-based approaches cannot go lower than 
the (4KB) page sized blocks as they hook into kernel 
paging; While LegoOS~\cite{legoos} and Kona~\cite{kona} 
escaped this fate through hardware modifications, 
other systems like AIFM~\cite{aifm} moved to user 
space implementations. Kona evaluates the effect of 
cache block size on the performance of Redis database 
and determines 1KB to be optimal, which is conveniently 
closer to the page size. However, Kona does not do any 
advanced prefetching (like Leap~\cite{leap}) which, in 
combination with smaller block sizes, may perform better 
than exploiting crude spatial locality with bigger blocks.
Block size also effects eviction policies like LRU (which
most systems use) as smaller blocks mean a larger number 
of blocks that need to be monitored for finding eviction
candidates. 

\vspace{3pt}
\noindent \uline{Prefetching.}
Prefetching remote data proactively can bring in 
correct pages into the cache and avoid cache misses 
in the critical path. Runtimes can use a transparent 
prefetcher that identifies access patterns and predict 
future accesses and/or they can provide prefetch calls 
in the interface that applications can inject in their 
code. While having a prefetcher is a more general 
solution, it has to balance between the accuracy of 
predictions and the (compute and memory) resources
it consumes (lower prefetching accuracy results in 
cache pollution and waste of cache and I/O bandwidth).
Leap~\cite{leap} is an advanced prefetcher for remote 
paging that monitors page faults and uses the faulted 
addresses to predict future pages. Even with coarse 
information like page faults, Leap was able to achieve 
1.5-2x improvement for different applications. 
Systems like AIFM~\cite{aifm} and Remote 
Regions~\cite{remregions} expose prefetch API 
providing applications the choice of implementing custom
prefetching such as the data structure-specific ones in AIFM. 

\vspace{3pt}
\noindent \uline{Cache Size.}
The bigger the size of the cache, the better; but the 
amount of local DRAM is limited (this limit is 
especially strict in hardware-disaggregated architecture 
where the amount of local DRAM is small and fixed). 
Even with the best prefetching and eviction policies, 
the cache size has to be enough to cover a minimum portion 
of the working set to avoid performance degradation and, 
in extreme cases, thrashing. A common chart we see in the 
evaluation of previous systems is to show the slowdown of 
an application against the local memory (or, cache size) as 
\% of either app's peak memory usage (which varies across 
apps) or an arbitrary local memory capacity, and compare 
it to other systems; the point being no one wants to pick 
a particular cache size but leave that option to the reader 
to trade it off with performance. Looking at variety of 
applications across papers, it seems like the performance 
degradation conforms to a hockey stick pattern, remaining 
graceful until some 25-50\% of the working 
set~\cite{netdisagg,kona,legoos} is local and degrading 
dramatically below that. None of these systems give 
an idea as to the one-size-fits-all limit for the absolute 
size of local DRAM though.

\subsubsection{Interface Overhead}
The interface itself can add some software overhead 
to the each memory access. Unlike virtual memory 
interfaces that use native load/store, user space 
interfaces involve either library calls or another 
level of pointer indirection (e.g., Java) that may add 
few cycles for each operation. For example, AIFM uses
C++ smart pointers and when compared to Fastswap that 
allows native pointers, it adds a marginal overhead 
that becomes evident when effects of their runtime and 
interconnect are minimized. (Fig 7~\cite{aifm}). 
Similarly, Kona~\cite{kona} that routes remote accesses
through the cache-coherent hardware that exposes remote 
memory as another physical DRAM whose accesses are 
slower compared to regular DRAM due to limited interconnect 
bandwidth. Kernel-based custom interfaces like 
LITE~\cite{literdma} introduce syscall in the access path 
that adds significant overhead. LITE however is a low-level 
interface that leaves caching to the application and 
such accesses can be made in the cache miss path.

\subsubsection{Remote Access Latency} 
Optimizing the actual latency in fetching remote data 
(i.e., the cache miss path) is important not only because 
caching cannot soften the performance impact if the miss 
latency is too high but also because it impacts tail-latencies 
that modern datacenter applications are sensitive to. 
This latency depends on both the implementation 
overheads on the client and the server side, and the choice of 
the interconnect itself. We discuss these factors below, 
in addition to the previous optimizations to reduce or 
hide this latency.

\vspace{3pt}
\noindent \uline{Network Interconnect.}
Gao et al.~\cite{netdisagg} analyzed the effect of increasing
remote access latency for various applications and arrive at 
4-5 $\mu$s upper bound to maintain performance. (It was, 
however, done on paging-based systems with default page 
replacement algorithms not optimized for remote memory, and 
hence should be taken as a rough estimate).
Such low latencies are now possible in tight-knit
local area networks like modern data center racks,  
where TCP/IP and, more recently, RDMA~\cite{rocev2} have
become standard transport options. 
RDMA has been the preferred target transport 
in most of the recent systems because it cuts down on the 
software stack on both sides and, more so, because it 
provides one-sided accesses that avoid remote CPU in 
critical path. For example, systems that~\cite{literdma,aifm}
explored the TCP/IP option reported up to 10 $\mu$s overhead 
compared to RDMA for remote operations. Technologies 
like Intel Omnipath~\cite{omnipath} and CCIX~\cite{ccix}
are expected to further slash the latency by bringing the NIC 
much closer to the CPU, once they are commercially available. 

\vspace{3pt}
\noindent \uline{Minimizing/avoiding software overheads.}
With respect to the remote access latency, the goal for the 
runtime is to get keep it as close as possible to the raw network 
latency and minimize any software overheads. The main software 
overhead comes from finding space for incoming remote data 
and deciding which data to evict to make that space 
(i.e., the typical cache miss handling), before returning to 
the application. 

Paging-based systems started with conventional disk-paging 
subsystem and over time improved it to target remote memory
where paging is more frequent and operates in microsecond 
timescales instead of milliseconds~\cite{Lim2012}. A common 
optimization is to decouple allocation and eviction by 
maintaining a free list for newly allocated or fetched-in 
data and moving eviction to the background and off the 
critical path~\cite{Lim2012,leap}. 
For example, Fastswap~\cite{fastswap} offloads  
memory reclamation to a dedicated CPU core so that the 
application CPU can return to user space and continue its 
execution. Another (defacto) optimization is to allocate 
local pages for newly allocated memory and keep it local until 
evicted (i.e., delayed write-back).
When bringing additional data along with the required data
(either because of prefetching or high data granularity to 
exploit locality), it is recommended to fetch the additional 
data separately outside the critical path~\cite{fastswap} to 
avoid head-of-the-line blocking.

Fundamentally though, as long as cache misses (page faults) are 
handled in software (the only available option with 
traditional hardware; although LegoOS, with hardware modifications, 
still takes the software route due to miss path complexity), 
the software miss path will inevitably involve overheads like 
context-switching, CPU cache pollution, etc. that add to the latency. 
Kona~\cite{kona}, for this reason, proposes new hardware 
primitives to offload this path to avoid above issues and also 
benefit from the resulting hardware speedup. The flip side, 
of course, is the complexity of implementing part of the 
runtime and networking stack in hardware.

\vspace{3pt}
\noindent \uline{Hiding the latency.}
From a throughput (CPU utilization) standpoint, latency 
can be hidden by switching out the current thread and 
running a different thread while the remote data is fetched.
Kernel thread context switches are on the order of 
microseconds and may not result in much savings 
but user space runtimes with light-weight 
threads can use this approach, like AIFM did.~\cite{aifm}.


\subsubsection{Reducing remote accesses with RPCs.}
Some access patterns like traversals may be ill-suited for remote 
fetching and cause too many cache misses no matter how good 
caching/prefetching is, and only viable option might be 
executing those operations closer to the memory using remote 
procedures (RPCs). 
To support this, systems may allow applications to register/invoke 
methods on the remote node, such as function shipping in 
FaRM~\cite{farm}, AIFM Remote Devices~\cite{aifm}, 
LITE RPC~\cite{literdma}, etc.
AIFM gets most of its performance benefits 
by sending such memory intensive operations of its data 
structures to the memory server. 
Semeru~\cite{semeru} does the same for 
garbage collecting the disaggregated Java Heap. Enabling this, 
however, depends on the assumed capabilities of the memory server.

\subsection{Memory Management}
Similar to virtual memory interface on traditional servers, a 
system for system for disaggregated memory needs to pool the 
memory on multiple servers/nodes and present a unified interface 
for allocating and accessing this memory by hiding away the 
underlying physical locations. It should be further responsible 
for tracking available memory across the nodes, map it to an  
application on allocation and fetch the data when required. 
Most systems~\cite{legoos,remregions,kona} employ a 
global memory manager that maintains this metadata and works 
with agents/daemons on the (memory) nodes to find space for 
new allocations. Compute nodes query the manager for new allocations 
and cache the mappings to directly go to the relevant memory node 
during access time. Under the hood, the system should aim for better 
memory efficiency while providing good data path performance and 
scalability. Below, we discuss few factors that affect these 
goals.

\subsubsection{Memory backing or complementing}
In a hardware disaggregated setting, all application memory is 
allocated in the disaggregated memory and local DRAM is very small
and only used as cache i.e, all the app memory is backed by 
disaggregated memory. In a software-disaggregated setup, 
a similar approach can be taken by reserving a small amount 
of local DRAM on each node for local workspace and the rest is 
pooled from which memory is transparently allocated to all the 
applications. However, this fails to exploit local affinity in such 
a non-uniform memory access setting so all systems proposed for 
traditional setting prioritize 
local allocations and only expose unutilized memory for shared 
cluster use. Such disaggregated pool can then be used for 
complementing (not backing) the local memory either by transparently 
expanding local address space (i.e., remote paging systems) or 
through other interfaces like mmapped files~\cite{remregions}.

\subsubsection{Memory tracking granularity.}
Just like paged memory, memory allocation and mapping is done 
is fixed-size units (slabs). Although extending page size (4KB) 
to remote memory would be a natural choice, 
it adds a significant 0.2\% space overhead 
for maintaining mapping metadata (e.g., 2 GB for a 1 TB 
region\cite{remregions}), so most systems end up using a much 
bigger slab size (128B~\cite{remregions} segments or ``multiple" 
pages~\cite{infiniswap,kona}). Local memory allocators can then 
manage these slabs for fine-grained allocations. 
However, bigger sizes may cause internal fragmentation or 
require contiguous segments at memory servers and hurt memory 
efficiency. None of the works, however, evaluate this aspect 
and slab sizes were arbitrary.

\subsubsection{Balancing memory usage across nodes.}
Remote memory used by an application is better if uniformly 
distributed across multiple memory nodes both to balance the 
access load and to minimize the performance impact in case of 
remote node failures. With a global memory manager, we can maintain 
the available memory across the memory nodes and properly direct 
memory allocation requests to both distribute memory footprint 
of application as well as balance overall memory usage across 
the memory nodes~\cite{legoos,remregions,kona}. However, both 
the overhead of constantly communicating with memory nodes 
and the centralized manager may present a scalability bottleneck.
Infiniswap~\cite{infiniswap} takes a decentralized approach 
where nodes requesting memory choose randomly from the list of 
available nodes. To help reduce imbalance, it opts for power of 
two choices~\cite{10.5555/924815} where instead of randomly 
picking one, two nodes are picked randomly and the one with the 
most available memory is chosen. However, it is unclear if this 
approach sustains the balance with increasing scale or over time.

%


\subsubsection{Memory reclamation on the server.}
As mentioned before, software-disaggregated systems need to 
balance available memory on each node between local and 
global memory (with priority for local use), and hence should 
should be amenable to reclamation to make space for expanding 
local usage. Infiniswap~\cite{infiniswap}, which 
uses remote memory as swap space, writes swapped out pages 
to both remote memory and disk, and hence can afford to drop the 
remote data during reclamation on a remote node. Reclamation is 
less of 
an issue in the hardware disaggregated setting but they may still 
need to swap out some memory to disk under extreme cluster 
memory pressure. An interesting, rather unexplored, question then
is which memory should be picked to drop/evict to minimize 
performance degradation. Traditional systems use LRU lists based 
on page access
information to swap out colder pages however such hotness information 
is either not available or, at best, distributed across 
compute nodes in a disaggregated system. 
Infiniswap~\cite{infiniswap} again uses power of choices to make an
informed guess (where the memory node queries a random subset of 
compute nodes for this hotness information and drop relatively 
colder pages) rather than randomly dropping any arbitrary pages.


\subsubsection{Summary}
Memory efficiency in general is not very well evaluated in any of 
the systems so far (only Infiniswap~\cite{infiniswap} had a chart 
showing cluster memory utilization over time), perhaps 
because of the focus on application 
performance which is the prime roadblock to feasible disaggregation.
Userspace runtimes completely punt on the memory management aspect 
and assume that required memory is pre-allocated~\cite{semeru} or 
limit themselves to working with a single memory server~\cite{aifm}. 
While they spotlighted the performance advantages of app integration, 
these systems most certainly need further work 
on the memory management side to be considered feasible for adoption. 

\subsection{Other Considerations}
There are other considerations that received less focus so far 
and were often overlooked in previous systems but needs to be 
worked on for adoption in the wild. As work on performance tuning 
saturates, we expect there to be more focus on these topics in future. 

\subsubsection{Fault Tolerance \& Reliability}
By distributing the memory across multiple servers/nodes, memory 
disaggregation expands the fault domain of an application 
and makes it more prone to failures. Our system should account
for this; we only need to worry about remote node 
failures as local node failures occur in traditional 
setup too and the goal is for our system to be as 
fault-tolerant as the traditional setup.

One option is in-memory replication~\cite{leap,kona}.
This, however, would consume at least twice as much memory,
wasting precious DRAM and may prove too high of a cost. 
LegoOS~\cite{legoos} only maintains the log on secondary 
replica to minimize the memory overhead risking higher 
recovery time after a failure. Another alternative 
is to write the remote pages/data to persistent storage (disk) 
in the background along with storing it in remote 
memory~\cite{infiniswap}. However, remote write load 
may be too high for disk to handle which may result in 
disk queue build-ups. Recovering from failure would also be 
slower but this may be tolerable if the failures are rare.
In both cases though, replication/writing to disk only happen 
during evictions which are generally off the critical path,
and hence won't directly affect application performance.
It is not clear, however, whether we can get away without 
any of these and just fail the applications using a memory
node when it crashes.

Since network is involved now, transient network failures
or congestion may affect remote accesses. Normally this only degrades 
performance due to timeouts/retries during page faults/misses
but does not affect fault tolerance. In Kona~\cite{kona} however,
remote accesses are served using special hardware through 
cache-coherent protocol that is sensitive to memory access 
latencies and may end up crashing the application.

\subsubsection{Network Efficiency}
Network bandwidth may become a limiting factor and so better 
network efficiency would keep the network less congested and 
latencies low. Amount of networked data is primarily affected 
by quality of caching (cache misses bring in the data) and 
the evictions (writing the dirty data back). Bigger cache
blocks, like the page size in most systems, will cause 
data amplification (by bringing in redundant data) and hurt 
network efficiency, especially during the write-backs where 
dirty data is usually only a small fraction of the block size. 
Paging approaches cannot track dirty data on finer granularity 
but Kona~\cite{kona} (with special hardware) and AIFM~\cite{aifm}
with userspace runtimes can, and avoid I/O amplification by just 
writing this data. For example, Kona uses a log to track multiple 
dirty cache lines and sends it to the memory node in batches.

\subsubsection{Security \& Isolation}
Kernel is (has to be) trusted, and as long as the 
runtime (and agents on all the components) work in 
kernel space, remote memory can be indirectly provided 
to application where these indirection mapping/translation 
is controlled by the kernel, providing a security 
isolation similar to that provided by the traditional 
virtual memory. All kernel-based systems provide this 
support to enable safe access to remote memory in presence
of multiple applications. User space runtimes cannot 
implement isolation and support sharing between 
applications as they are restricted to a single 
application. Such runtimes may need to 
bank on kernel-based, lower-level interfaces like 
Remote Regions~\cite{remregions} or LITE~\cite{literdma}
for proper isolation in presence of multiple applications 
using such runtimes. Even if an attacker does not 
control these components, disaggregated memory increases the 
attack surface and may enable side-channel attacks such as  
Pythia~\cite{Pythia}, so more work on security is certainly 
needed. 




\section{DISCUSSION}
\label{sec:discussion}
In this section, we first discuss some open questions regarding 
the design that haven't been categorically answered so far. 
We then talk about a few potential directions for future 
exploration.

\subsection{Open Questions}
\vspace{3pt}
\noindent \uline{Transparency vs. App Integration}
Transparency is important for adoption so, as we've seen, 
most previous systems opt for it. However, as 
language-based runtimes have shown, exploiting 
application hints/semantics for optimizing runtime can be
a key factor in cutting down the performance impact.
Perhaps there is a middle-ground to be found here 
where a system provides both options and the choice is left
to the application to either run without changes or 
provide hints for better performance. For example, 
a kernel-based virtual memory interface that also exposes 
a custom API through 
ioctl (like LITE~\cite{literdma}) to, for example, supply 
application-specific prefetching or call an RPC.

\vspace{3pt}
\noindent \uline{The right amount of memory sharing (and consistency).}
Systems so far have either completely avoided any explicit 
memory sharing across servers or provided only coarse-grained 
sharing with no consistency guarantees. 
If sharing is to be supported for transparent virtual 
memory interface, it should provide same semantics as current 
cache coherence protocols on the entire address space but 
providing such strong consistency model across the network 
is still infeasible (even the recent RDMA-based DSMs~\cite{farm,gam} 
do not attempt it); none of the major (transparent) 
disaggregation systems~\cite{infiniswap,legoos} provided this 
support, perhaps for this reason. Non-transparent ones like 
Remote Regions~\cite{remregions} and LITE~\cite{literdma} provide
support for mapping the same region in multiple servers along with
basic synchronization support like barriers and mutexes for 
synchronizing their accesses, but no consistency on reads/writes. 
Sharing generally disallows caching and delayed write-backs,
both of which are critical to making performance of remote memory 
feasible for applications, as we have seen in 
section~\ref{sec:performance}. Aguilera et al.~\cite{Aguilera2017}
mooted the idea of non-simultaneous sharing where at any time, 
memory is accessed exclusively by a single host, but across 
time it can be accessed by many; which is enough for some common 
distributed workloads like MapReduce.

\vspace{3pt}
\noindent \uline{The extent of memory-side compute.}
Most previous systems differ significantly on the 
amount of remote/near-memory compute capabilities they expect 
of the memory server. At the bare minimum, the memory server 
should be able to serve remote memory allocations and accesses while 
managing its local memory. On top of that, some systems offload 
other runtime tasks such as address mapping, replication and 
persisting to disk~\cite{legoos}, and even garbage 
collection~\cite{semeru} to the memory servers. 
Others~\cite{aifm,literdma} go even further and introduce 
RPCs that applications can register/run on the memory server, 
which can be of arbitrary complexity but help with performance. 
However, this may be a slippery slope as it adds 
complexity to the memory server that we do not want, at least in 
the hardware disaggregated setting where there is limited compute 
on memory nodes. Even in the traditional setup, we are still 
looking to decouple CPU from memory, and  increasing compute 
complexity on the remote memory side does not help with 
disaggregation and only complicates resource accounting. 
Perhaps a middle ground is to find a set of hardware-friendly 
primitives 
(as in~\cite{Aguilera2019}) for common remote side operations and 
expose only these set as RPCs through platforms like 
Storm~\cite{storm} and Strom~\cite{strom} that enable  
implementing these RPCs on the remote NICs. 

\vspace{3pt}
\noindent \uline{The right benchmark to evaluate with.}
A disaggregated system is expected to be general and support
a wide variety of applications, not just interface-wise but 
also w.r.t. the performance impact. As many recent 
systems show in their evaluation~\cite{netdisagg,legoos,fastswap}, 
performance degradation can vary dramatically across 
applications depending on the amount of local memory.
Similarly, different applications react differently to 
perfomance optimizations like caching and prefetching,
based on their memory access and locality patterns. 
Given this heterogeneity, the benchmark for evaluating a 
diaggregated memory system should include a variety of 
applications ranging from compute to memory intensive ones. 
All recent systems, however, chose their own set of custom 
benchmarking applications and very few of them intersect 
making the comparison across the spectrum infeasible as 
the chosen benchmark may favor their own systems. 
It is, therefore, preferable to have a standard benchmark.

\subsection{More Opportunities for Future}
\vspace{3pt}
\noindent \uline{Disaggregated Memory for VMs and beyond.}
Most modern workloads run on some virtualization platform 
or the other (VMs, Containers, Lambdas, etc.)~\cite{Aguilera2017}, 
so there is a need to extend disaggregated memory to such 
platforms for wider adoption. The hypervisor can take the place 
of the kernel in implementing runtime and managing/providing 
the diaggregated memory. However, VMs, unlike processes, come 
with strict SLOs and with an expectation of performance isolation,
so the runtime needs to account for this while managing memory 
(e.g., proper memory accouting, maintaining separate caches, etc.). 
Also, the gap between the application and the runtime is further 
widened making it harder to get app-specific information.

\vspace{3pt}
\noindent \uline{Job scheduling on disaggregated systems.}
Ideally, job scheduling based on its CPU and memory requirements 
should be simple on a disaggregated system because as long 
as the cluster has enough memory for a job, it can be placed  
anywhere. However, non-uniform memory access between local 
and remote memory means that cluster throughput would be 
higher if total number of remote accesses, which can depend
on job placement, is minimized. For example, given similar 
compute requirements, jobs with large working memory are   
better paired with the ones with smaller working sets on the 
individual compute nodes to balance out contention for the 
local DRAM (cache). Similarly, different jobs react differently 
to reduced local memory (cache) size which is another aspect on 
which jobs can be balanced to prioritize the sensitive ones.
This line of work can build on far-memory aware schedulers 
like Fastswap~\cite{fastswap}.

\vspace{3pt}
\noindent \uline{Exploiting other new trends 
in the datacenters.}
With datacenters increasingly employing heterogeneous 
hardware, disaggregated memory should be made accessible 
to these devices as well. This raises the question of 
the best interface for accessing memory from such  
custom devices like GPUs or FPGAs, and how feasible it is  
for implementation on these devices. For example, 
traditional paging-based interfaces cannot be 
implemented on custom hardware that lacks the TLB and MMU 
hardware (LegoOS~\cite{legoos} moves the translation hardware 
to memory nodes making it friendlier for such hardware). 
Another related trend is the 
adoption of programmable networking hardware like 
NICs and switches. Switches have been used to 
accelerate applications through in-switch caching, load
balancing, etc. Similar acceleration opporunities may be found 
in case of diaggregated memory systems for global tasks like 
memory management and load balancing in the critical path.


%

\bibliographystyle{ACM-Reference-Format}
\bibliography{references}


\begin{thebibliography}{32}


\ifx \showCODEN    \undefined \def \showCODEN     #1{\unskip}     \fi
\ifx \showDOI      \undefined \def \showDOI       #1{#1}\fi
\ifx \showISBNx    \undefined \def \showISBNx     #1{\unskip}     \fi
\ifx \showISBNxiii \undefined \def \showISBNxiii  #1{\unskip}     \fi
\ifx \showISSN     \undefined \def \showISSN      #1{\unskip}     \fi
\ifx \showLCCN     \undefined \def \showLCCN      #1{\unskip}     \fi
\ifx \shownote     \undefined \def \shownote      #1{#1}          \fi
\ifx \showarticletitle \undefined \def \showarticletitle #1{#1}   \fi
\ifx \showURL      \undefined \def \showURL       {\relax}        \fi
\providecommand\bibfield[2]{#2}
\providecommand\bibinfo[2]{#2}
\providecommand\natexlab[1]{#1}
\providecommand\showeprint[2][]{arXiv:#2}

\bibitem[\protect\citeauthoryear{??}{cci}{2021}]%
        {ccix}
 \bibinfo{year}{2021}\natexlab{}.
\newblock \bibinfo{title}{CCIX: cache coherent interconnect for accelerators}.
\newblock \bibinfo{howpublished}{https://www.ccixconsortium.com/}.
\newblock


\bibitem[\protect\citeauthoryear{??}{omn}{2021}]%
        {omnipath}
 \bibinfo{year}{2021}\natexlab{}.
\newblock \bibinfo{title}{Intel OmniPath Architecture}.
\newblock
  \bibinfo{howpublished}{https://www.intel.com/content/www/us/en/high-performance-computing-fabrics/omni-path-driving-exascale-computing.html}.
\newblock


\bibitem[\protect\citeauthoryear{Aguilera, Amit, Calciu, Deguillard, Gandhi,
  Subrahmanyam, Suresh, Tati, Venkatasubramanian, and Wei}{Aguilera
  et~al\mbox{.}}{2017}]%
        {Aguilera2017}
\bibfield{author}{\bibinfo{person}{Marcos~K. Aguilera}, \bibinfo{person}{Nadav
  Amit}, \bibinfo{person}{Irina Calciu}, \bibinfo{person}{Xavier Deguillard},
  \bibinfo{person}{Jayneel Gandhi}, \bibinfo{person}{Pratap Subrahmanyam},
  \bibinfo{person}{Lalith Suresh}, \bibinfo{person}{Kiran Tati},
  \bibinfo{person}{Rajesh Venkatasubramanian}, {and} \bibinfo{person}{Michael
  Wei}.} \bibinfo{year}{2017}\natexlab{}.
\newblock \showarticletitle{{Remote memory in the age of fast networks}}.
\newblock \bibinfo{journal}{\emph{SoCC 2017 - Proceedings of the 2017 Symposium
  on Cloud Computing}} (\bibinfo{year}{2017}), \bibinfo{pages}{121--127}.
\newblock
\showISBNx{9781450350280}
\urldef\tempurl%
\url{https://doi.org/10.1145/3127479.3131612}
\showDOI{\tempurl}


\bibitem[\protect\citeauthoryear{Aguilera, Amit, Calciu, Deguillard, Novakovic,
  Ramanathan, Suresh, Tati, Venkatasubramanian, and Wei}{Aguilera
  et~al\mbox{.}}{2020}]%
        {remregions}
\bibfield{author}{\bibinfo{person}{Marcos~K. Aguilera}, \bibinfo{person}{Nadav
  Amit}, \bibinfo{person}{Irina Calciu}, \bibinfo{person}{Xavier Deguillard},
  \bibinfo{person}{Jayneel Gandhi~Stanko Novakovic}, \bibinfo{person}{Arun
  Ramanathan}, \bibinfo{person}{Pratap Subrahmanyam~Lalith Suresh},
  \bibinfo{person}{Kiran Tati}, \bibinfo{person}{Rajesh Venkatasubramanian},
  {and} \bibinfo{person}{Michael Wei}.} \bibinfo{year}{2020}\natexlab{}.
\newblock \showarticletitle{{Remote regions: A simple abstraction for remote
  memory}}.
\newblock \bibinfo{journal}{\emph{Proceedings of the 2018 USENIX Annual
  Technical Conference, USENIX ATC 2018}} (\bibinfo{year}{2020}),
  \bibinfo{pages}{775--787}.
\newblock
\showISBNx{9781939133021}


\bibitem[\protect\citeauthoryear{Aguilera, Keeton, Novakovic, and
  Singhal}{Aguilera et~al\mbox{.}}{2019}]%
        {Aguilera2019}
\bibfield{author}{\bibinfo{person}{Marcos~K. Aguilera},
  \bibinfo{person}{Kimberly Keeton}, \bibinfo{person}{Stanko Novakovic}, {and}
  \bibinfo{person}{Sharad Singhal}.} \bibinfo{year}{2019}\natexlab{}.
\newblock \showarticletitle{{Designing Far Memory Data Structures: Think
  Outside the Box}}.
\newblock \bibinfo{journal}{\emph{Proceedings of the Workshop on Hot Topics in
  Operating Systems, HotOS 2019}} (\bibinfo{year}{2019}),
  \bibinfo{pages}{120--126}.
\newblock
\showISBNx{9781450367271}
\urldef\tempurl%
\url{https://doi.org/10.1145/3317550.3321433}
\showDOI{\tempurl}


\bibitem[\protect\citeauthoryear{{Al Maruf} and Chowdhury}{{Al Maruf} and
  Chowdhury}{2019}]%
        {leap}
\bibfield{author}{\bibinfo{person}{Hassan {Al Maruf}} {and}
  \bibinfo{person}{Mosharaf Chowdhury}.} \bibinfo{year}{2019}\natexlab{}.
\newblock \bibinfo{booktitle}{\emph{{Effectively Prefetching Remote Memory with
  Leap}}}.
\newblock
\showISBNx{9781939133144}
\showISSN{23318422}
\urldef\tempurl%
\url{https://www.usenix.org/conference/atc20/presentation/al-maruf}
\showURL{%
\tempurl}


\bibitem[\protect\citeauthoryear{Amaro, Branner-Augmon, Luo, Ousterhout,
  Aguilera, Panda, Ratnasamy, and Shenker}{Amaro et~al\mbox{.}}{2020}]%
        {fastswap}
\bibfield{author}{\bibinfo{person}{Emmanuel Amaro},
  \bibinfo{person}{Christopher Branner-Augmon}, \bibinfo{person}{Zhihong Luo},
  \bibinfo{person}{Amy Ousterhout}, \bibinfo{person}{Marcos~K. Aguilera},
  \bibinfo{person}{Aurojit Panda}, \bibinfo{person}{Sylvia Ratnasamy}, {and}
  \bibinfo{person}{Scott Shenker}.} \bibinfo{year}{2020}\natexlab{}.
\newblock \showarticletitle{{Can far memory improve job throughput?}}. In
  \bibinfo{booktitle}{\emph{Proceedings of the 15th European Conference on
  Computer Systems, EuroSys 2020}}. \bibinfo{publisher}{ACM},
  \bibinfo{pages}{16}.
\newblock
\showISBNx{9781450368827}
\urldef\tempurl%
\url{https://doi.org/10.1145/3342195.3387522}
\showDOI{\tempurl}


\bibitem[\protect\citeauthoryear{Cai, Guo, Zhang, Agrawal, Chenz, Ooi, Tan,
  Teo, and Wang}{Cai et~al\mbox{.}}{2018}]%
        {gam}
\bibfield{author}{\bibinfo{person}{Qingchao Cai}, \bibinfo{person}{Wentian
  Guo}, \bibinfo{person}{Hao Zhang}, \bibinfo{person}{Divyakant Agrawal},
  \bibinfo{person}{Gang Chenz}, \bibinfo{person}{Beng~Chin Ooi},
  \bibinfo{person}{Kian~Lee Tan}, \bibinfo{person}{Yong~Meng Teo}, {and}
  \bibinfo{person}{Sheng Wang}.} \bibinfo{year}{2018}\natexlab{}.
\newblock \showarticletitle{{Efficient distributed memory management with RDMA
  and caching}}.
\newblock \bibinfo{journal}{\emph{Proceedings of the VLDB Endowment}}
  \bibinfo{volume}{11}, \bibinfo{number}{11} (\bibinfo{year}{2018}),
  \bibinfo{pages}{1604--1617}.
\newblock
\showISSN{21508097}
\urldef\tempurl%
\url{https://doi.org/10.14778/3236187.3236209}
\showDOI{\tempurl}


\bibitem[\protect\citeauthoryear{Calciu, Imran, Puddu, Kashyap, Maruf, Mutlu,
  and Kolli}{Calciu et~al\mbox{.}}{2021}]%
        {kona}
\bibfield{author}{\bibinfo{person}{Irina Calciu}, \bibinfo{person}{M.~Talha
  Imran}, \bibinfo{person}{Ivan Puddu}, \bibinfo{person}{Sanidhya Kashyap},
  \bibinfo{person}{Hasan~Al Maruf}, \bibinfo{person}{Onur Mutlu}, {and}
  \bibinfo{person}{Aasheesh Kolli}.} \bibinfo{year}{2021}\natexlab{}.
\newblock \showarticletitle{{Rethinking software runtimes for disaggregated
  memory}}. In \bibinfo{booktitle}{\emph{ASPLOS}}. \bibinfo{pages}{79--92}.
\newblock
\showISBNx{9781450383172}
\urldef\tempurl%
\url{https://doi.org/10.1145/3445814.3446713}
\showDOI{\tempurl}


\bibitem[\protect\citeauthoryear{Dragojevi{\'{c}}, Narayanan, Hodson, and
  Castro}{Dragojevi{\'{c}} et~al\mbox{.}}{2014}]%
        {farm}
\bibfield{author}{\bibinfo{person}{Aleksandar Dragojevi{\'{c}}},
  \bibinfo{person}{Dushyanth Narayanan}, \bibinfo{person}{Orion Hodson}, {and}
  \bibinfo{person}{Miguel Castro}.} \bibinfo{year}{2014}\natexlab{}.
\newblock \showarticletitle{{FaRM: Fast remote memory}}.
\newblock \bibinfo{journal}{\emph{Proceedings of the 11th USENIX Symposium on
  Networked Systems Design and Implementation, NSDI 2014}}
  (\bibinfo{year}{2014}), \bibinfo{pages}{401--414}.
\newblock
\showISBNx{9781931971096}


\bibitem[\protect\citeauthoryear{Dwarkadas, Hardavellas, Kontothanassis,
  Nikhil, and Stets}{Dwarkadas et~al\mbox{.}}{1999}]%
        {cashmere}
\bibfield{author}{\bibinfo{person}{Sandhya Dwarkadas},
  \bibinfo{person}{Nikolaos Hardavellas}, \bibinfo{person}{Leonidas
  Kontothanassis}, \bibinfo{person}{Rishiyur Nikhil}, {and}
  \bibinfo{person}{Robert Stets}.} \bibinfo{year}{1999}\natexlab{}.
\newblock \showarticletitle{{Cashmere-VLM: Remote memory paging for software
  distributed shared memory}}.
\newblock \bibinfo{journal}{\emph{Proceedings of the International Parallel
  Processing Symposium, IPPS}} (\bibinfo{year}{1999}),
  \bibinfo{pages}{153--159}.
\newblock
\showISSN{10637133}
\urldef\tempurl%
\url{https://doi.org/10.1109/ipps.1999.760451}
\showDOI{\tempurl}


\bibitem[\protect\citeauthoryear{Feeley, Morgan, Pighin, Karlin, Levy, and
  Thekkath}{Feeley et~al\mbox{.}}{1995}]%
        {gms}
\bibfield{author}{\bibinfo{person}{M.~J. Feeley}, \bibinfo{person}{W.~E.
  Morgan}, \bibinfo{person}{E.~P. Pighin}, \bibinfo{person}{A.~R. Karlin},
  \bibinfo{person}{H.~M. Levy}, {and} \bibinfo{person}{C.~A. Thekkath}.}
  \bibinfo{year}{1995}\natexlab{}.
\newblock \showarticletitle{Implementing Global Memory Management in a
  Workstation Cluster}. In \bibinfo{booktitle}{\emph{Proceedings of the
  Fifteenth ACM Symposium on Operating Systems Principles}} (Copper Mountain,
  Colorado, USA) \emph{(\bibinfo{series}{SOSP '95})}.
  \bibinfo{publisher}{Association for Computing Machinery},
  \bibinfo{address}{New York, NY, USA}, \bibinfo{pages}{201–212}.
\newblock
\showISBNx{0897917154}
\urldef\tempurl%
\url{https://doi.org/10.1145/224056.224072}
\showDOI{\tempurl}


\bibitem[\protect\citeauthoryear{Gao, Narayan, Karandikar, Carreira, Han,
  Agarwal, Ratnasamy, and Shenker}{Gao et~al\mbox{.}}{2016}]%
        {netdisagg}
\bibfield{author}{\bibinfo{person}{Peter~X. Gao}, \bibinfo{person}{Akshay
  Narayan}, \bibinfo{person}{Sagar Karandikar}, \bibinfo{person}{Joao
  Carreira}, \bibinfo{person}{Sangjin Han}, \bibinfo{person}{Rachit Agarwal},
  \bibinfo{person}{Sylvia Ratnasamy}, {and} \bibinfo{person}{Scott Shenker}.}
  \bibinfo{year}{2016}\natexlab{}.
\newblock \showarticletitle{Network Requirements for Resource Disaggregation}.
  In \bibinfo{booktitle}{\emph{12th {USENIX} Symposium on Operating Systems
  Design and Implementation ({OSDI} 16)}}. \bibinfo{publisher}{{USENIX}
  Association}, \bibinfo{address}{Savannah, GA}, \bibinfo{pages}{249--264}.
\newblock
\showISBNx{978-1-931971-33-1}
\urldef\tempurl%
\url{https://www.usenix.org/conference/osdi16/technical-sessions/presentation/gao}
\showURL{%
\tempurl}


\bibitem[\protect\citeauthoryear{Gu, Lee, Zhang, Chowdhury, and Shin}{Gu
  et~al\mbox{.}}{2017}]%
        {infiniswap}
\bibfield{author}{\bibinfo{person}{Juncheng Gu}, \bibinfo{person}{Youngmoon
  Lee}, \bibinfo{person}{Yiwen Zhang}, \bibinfo{person}{Mosharaf Chowdhury},
  {and} \bibinfo{person}{Kang~G. Shin}.} \bibinfo{year}{2017}\natexlab{}.
\newblock \bibinfo{booktitle}{\emph{{Efficient memory disaggregation with
  Infiniswap}}}.
\newblock 649--667 pages.
\newblock
\showISBNx{9781931971379}
\urldef\tempurl%
\url{https://www.usenix.org/conference/nsdi17/technical-sessions/presentation/gu}
\showURL{%
\tempurl}


\bibitem[\protect\citeauthoryear{Kalia, Kaminsky, and Andersen}{Kalia
  et~al\mbox{.}}{2015}]%
        {Kalia2015}
\bibfield{author}{\bibinfo{person}{Anuj Kalia}, \bibinfo{person}{Michael
  Kaminsky}, {and} \bibinfo{person}{David~G. Andersen}.}
  \bibinfo{year}{2015}\natexlab{}.
\newblock \showarticletitle{{Using RDMA efficiently for key-value services}}.
\newblock \bibinfo{journal}{\emph{Computer Communication Review}}
  \bibinfo{volume}{44}, \bibinfo{number}{4} (\bibinfo{year}{2015}),
  \bibinfo{pages}{295--306}.
\newblock
\showISBNx{9781450328364}
\showISSN{19435819}
\urldef\tempurl%
\url{https://doi.org/10.1145/2619239.2626299}
\showDOI{\tempurl}


\bibitem[\protect\citeauthoryear{Keleher, Cox, Dwarkadas, and
  Zwaenepoel}{Keleher et~al\mbox{.}}{1994}]%
        {treadmarks}
\bibfield{author}{\bibinfo{person}{Pete Keleher}, \bibinfo{person}{Alan~L Cox},
  \bibinfo{person}{Sandhya Dwarkadas}, {and} \bibinfo{person}{Willy
  Zwaenepoel}.} \bibinfo{year}{1994}\natexlab{}.
\newblock \showarticletitle{{TreadMarks: Distributed Shared Memory on Standard
  Workstations and Operating Systems}}.
\newblock  (\bibinfo{year}{1994}).
\newblock


\bibitem[\protect\citeauthoryear{Lagar-Cavilla, Ahn, Souhlal, Agarwal, Burny,
  Butt, Chang, Chaugule, Deng, Shahid, Thelen, Yurtsever, Zhao, and
  Ranganathan}{Lagar-Cavilla et~al\mbox{.}}{2019}]%
        {zswap}
\bibfield{author}{\bibinfo{person}{Andres Lagar-Cavilla},
  \bibinfo{person}{Junwhan Ahn}, \bibinfo{person}{Suleiman Souhlal},
  \bibinfo{person}{Neha Agarwal}, \bibinfo{person}{Radoslaw Burny},
  \bibinfo{person}{Shakeel Butt}, \bibinfo{person}{Jichuan Chang},
  \bibinfo{person}{Ashwin Chaugule}, \bibinfo{person}{Nan Deng},
  \bibinfo{person}{Junaid Shahid}, \bibinfo{person}{Greg Thelen},
  \bibinfo{person}{Kamil~Adam Yurtsever}, \bibinfo{person}{Yu Zhao}, {and}
  \bibinfo{person}{Parthasarathy Ranganathan}.}
  \bibinfo{year}{2019}\natexlab{}.
\newblock \showarticletitle{{Software-Defined Far Memory in Warehouse-Scale
  Computers}}.
\newblock \bibinfo{journal}{\emph{International Conference on Architectural
  Support for Programming Languages and Operating Systems - ASPLOS}}
  (\bibinfo{year}{2019}), \bibinfo{pages}{317--330}.
\newblock
\showISBNx{9781450362405}
\urldef\tempurl%
\url{https://doi.org/10.1145/3297858.3304053}
\showDOI{\tempurl}


\bibitem[\protect\citeauthoryear{Li and Hudak}{Li and Hudak}{1989}]%
        {dsm1}
\bibfield{author}{\bibinfo{person}{Kai Li} {and} \bibinfo{person}{Paul Hudak}.}
  \bibinfo{year}{1989}\natexlab{}.
\newblock \showarticletitle{{Memory Coherence in Shared Virtual Memory
  Systems}}.
\newblock \bibinfo{journal}{\emph{ACM Transactions on Computer Systems (TOCS)}}
  \bibinfo{volume}{7}, \bibinfo{number}{4} (\bibinfo{year}{1989}),
  \bibinfo{pages}{321--359}.
\newblock
\showISSN{15577333}
\urldef\tempurl%
\url{https://doi.org/10.1145/75104.75105}
\showDOI{\tempurl}


\bibitem[\protect\citeauthoryear{Lim, Chang, Mudge, Ranganathan, Reinhardt, and
  Wenisch}{Lim et~al\mbox{.}}{2009}]%
        {bladedisagg1}
\bibfield{author}{\bibinfo{person}{Kevin Lim}, \bibinfo{person}{Jichuan Chang},
  \bibinfo{person}{Trevor Mudge}, \bibinfo{person}{Parthasarathy Ranganathan},
  \bibinfo{person}{Steven~K. Reinhardt}, {and} \bibinfo{person}{Thomas~F.
  Wenisch}.} \bibinfo{year}{2009}\natexlab{}.
\newblock \showarticletitle{{Disaggregated memory for expansion and sharing in
  blade servers}}. In \bibinfo{booktitle}{\emph{Proceedings - International
  Symposium on Computer Architecture}}. \bibinfo{pages}{267--278}.
\newblock
\showISBNx{9781605585260}
\showISSN{10636897}
\urldef\tempurl%
\url{https://doi.org/10.1145/1555754.1555789}
\showDOI{\tempurl}


\bibitem[\protect\citeauthoryear{Lim, Turner, Santos, AuYoung, Chang,
  Ranganathan, and Wenisch}{Lim et~al\mbox{.}}{2012}]%
        {Lim2012}
\bibfield{author}{\bibinfo{person}{Kevin Lim}, \bibinfo{person}{Yoshio Turner},
  \bibinfo{person}{Jose~Renato Santos}, \bibinfo{person}{Alvin AuYoung},
  \bibinfo{person}{Jichuan Chang}, \bibinfo{person}{Parthasarathy Ranganathan},
  {and} \bibinfo{person}{Thomas~F. Wenisch}.} \bibinfo{year}{2012}\natexlab{}.
\newblock \bibinfo{booktitle}{\emph{{System-level implications of disaggregated
  memory}}}.
\newblock \bibinfo{type}{{T}echnical {R}eport}. \bibinfo{pages}{1--12} pages.
\newblock
\urldef\tempurl%
\url{https://doi.org/10.1109/hpca.2012.6168955}
\showDOI{\tempurl}


\bibitem[\protect\citeauthoryear{Mitzenmacher and Sinclair}{Mitzenmacher and
  Sinclair}{1996}]%
        {10.5555/924815}
\bibfield{author}{\bibinfo{person}{Michael~David Mitzenmacher} {and}
  \bibinfo{person}{Alistair Sinclair}.} \bibinfo{year}{1996}\natexlab{}.
\newblock \emph{\bibinfo{title}{The Power of Two Choices in Randomized Load
  Balancing}}.
\newblock \bibinfo{thesistype}{Ph.D. Dissertation}.
\newblock
\showISBNx{0591320916}
\newblock
\shownote{AAI9723118.}


\bibitem[\protect\citeauthoryear{Novakovi{\'{c}}, Daglis, Bugnion, Falsafi, and
  Grot}{Novakovi{\'{c}} et~al\mbox{.}}{2014}]%
        {sonuma}
\bibfield{author}{\bibinfo{person}{Stanko Novakovi{\'{c}}},
  \bibinfo{person}{Alexandros Daglis}, \bibinfo{person}{Edouard Bugnion},
  \bibinfo{person}{Babak Falsafi}, {and} \bibinfo{person}{Boris Grot}.}
  \bibinfo{year}{2014}\natexlab{}.
\newblock \showarticletitle{{Scale-Out NUMA}}.
\newblock \bibinfo{journal}{\emph{International Conference on Architectural
  Support for Programming Languages and Operating Systems - ASPLOS}}
  (\bibinfo{year}{2014}), \bibinfo{pages}{3--17}.
\newblock
\showISBNx{9781450323055}
\urldef\tempurl%
\url{https://doi.org/10.1145/2541940.2541965}
\showDOI{\tempurl}


\bibitem[\protect\citeauthoryear{Novakovic, Daglis, Bugnion, Falsafi, and
  Grot}{Novakovic et~al\mbox{.}}{2016}]%
        {Novakovic2016}
\bibfield{author}{\bibinfo{person}{Stanko Novakovic},
  \bibinfo{person}{Alexandros Daglis}, \bibinfo{person}{Edouard Bugnion},
  \bibinfo{person}{Babak Falsafi}, {and} \bibinfo{person}{Boris Grot}.}
  \bibinfo{year}{2016}\natexlab{}.
\newblock \showarticletitle{{The case for rackout: Scalable data serving using
  rack-scale systems}}. In \bibinfo{booktitle}{\emph{Proceedings of the 7th ACM
  Symposium on Cloud Computing, SoCC 2016}}. \bibinfo{pages}{182--195}.
\newblock
\showISBNx{9781450345255}
\urldef\tempurl%
\url{https://doi.org/10.1145/2987550.2987577}
\showDOI{\tempurl}


\bibitem[\protect\citeauthoryear{Novakovic, Shan, Kolli, Cui, Zhang, Eran,
  Pismenny, Liss, Wei, Tsafrir, and Aguilera}{Novakovic et~al\mbox{.}}{2019}]%
        {storm}
\bibfield{author}{\bibinfo{person}{Stanko Novakovic}, \bibinfo{person}{Yizhou
  Shan}, \bibinfo{person}{Aasheesh Kolli}, \bibinfo{person}{Michael Cui},
  \bibinfo{person}{Yiying Zhang}, \bibinfo{person}{Haggai Eran},
  \bibinfo{person}{Boris Pismenny}, \bibinfo{person}{Liran Liss},
  \bibinfo{person}{Michael Wei}, \bibinfo{person}{Dan Tsafrir}, {and}
  \bibinfo{person}{Marcos Aguilera}.} \bibinfo{year}{2019}\natexlab{}.
\newblock \showarticletitle{{StoRM: A fast transactional dataplane for remote
  data structures}}.
\newblock \bibinfo{journal}{\emph{SYSTOR 2019 - Proceedings of the 12th ACM
  International Systems and Storage Conference}} (\bibinfo{year}{2019}),
  \bibinfo{pages}{97--108}.
\newblock
\showISBNx{9781450367493}
\urldef\tempurl%
\url{https://doi.org/10.1145/3319647.3325827}
\showDOI{\tempurl}
\showeprint[arxiv]{1902.02411}


\bibitem[\protect\citeauthoryear{Ousterhout, Agrawal, Erickson, Kozyrakis,
  Leverich, Mazi{\`{e}}res, Mitra, Narayanan, Parulkar, Rosenblum, Rumble,
  Stratmann, and Stutsman}{Ousterhout et~al\mbox{.}}{2009}]%
        {Ousterhout2010}
\bibfield{author}{\bibinfo{person}{John Ousterhout}, \bibinfo{person}{Parag
  Agrawal}, \bibinfo{person}{David Erickson}, \bibinfo{person}{Christos
  Kozyrakis}, \bibinfo{person}{Jacob Leverich}, \bibinfo{person}{David
  Mazi{\`{e}}res}, \bibinfo{person}{Subhasish Mitra}, \bibinfo{person}{Aravind
  Narayanan}, \bibinfo{person}{Guru Parulkar}, \bibinfo{person}{Mendel
  Rosenblum}, \bibinfo{person}{Stephen~M Rumble}, \bibinfo{person}{Eric
  Stratmann}, {and} \bibinfo{person}{Ryan Stutsman}.}
  \bibinfo{year}{2009}\natexlab{}.
\newblock \showarticletitle{{The Case for RAMClouds: Scalable High-Performance
  Storage Entirely in DRAM}}.
\newblock \bibinfo{journal}{\emph{Stanfordedu}}  \bibinfo{volume}{43}
  (\bibinfo{year}{2009}), \bibinfo{pages}{1--14}.
\newblock
\urldef\tempurl%
\url{http://www.stanford.edu/$\sim$ouster/cgi-bin/papers/ramcloud.pdf}
\showURL{%
\tempurl}


\bibitem[\protect\citeauthoryear{Ruan, Schwarzkopf, Aguilera, and Belay}{Ruan
  et~al\mbox{.}}{2020}]%
        {aifm}
\bibfield{author}{\bibinfo{person}{Zhenyuan Ruan}, \bibinfo{person}{Malte
  Schwarzkopf}, \bibinfo{person}{Marcos~K. Aguilera}, {and}
  \bibinfo{person}{Adam Belay}.} \bibinfo{year}{2020}\natexlab{}.
\newblock \showarticletitle{{AIFM: High-Performance, application-integrated far
  memory}}.
\newblock \bibinfo{journal}{\emph{Proceedings of the 14th USENIX Symposium on
  Operating Systems Design and Implementation, OSDI 2020}}
  (\bibinfo{year}{2020}), \bibinfo{pages}{315--332}.
\newblock
\showISBNx{9781939133199}


\bibitem[\protect\citeauthoryear{Shan, Huang, Chen, and Zhang}{Shan
  et~al\mbox{.}}{2007}]%
        {legoos}
\bibfield{author}{\bibinfo{person}{Yizhou Shan}, \bibinfo{person}{Yutong
  Huang}, \bibinfo{person}{Yilun Chen}, {and} \bibinfo{person}{Yiying Zhang}.}
  \bibinfo{year}{2007}\natexlab{}.
\newblock \showarticletitle{{Legoos: A disseminated, distributed OS for
  hardware resource disaggregation}}.
\newblock \bibinfo{journal}{\emph{Proceedings of the 13th USENIX Symposium on
  Operating Systems Design and Implementation, OSDI 2018}}
  (\bibinfo{year}{2007}), \bibinfo{pages}{69--87}.
\newblock
\showISBNx{9781939133083}
\urldef\tempurl%
\url{https://www.usenix.org/conference/osdi18/presentation/shan}
\showURL{%
\tempurl}


\bibitem[\protect\citeauthoryear{Sidler, Wang, Chiosa, Kulkarni, and
  Alonso}{Sidler et~al\mbox{.}}{2020}]%
        {strom}
\bibfield{author}{\bibinfo{person}{David Sidler}, \bibinfo{person}{Zeke Wang},
  \bibinfo{person}{Monica Chiosa}, \bibinfo{person}{Amit Kulkarni}, {and}
  \bibinfo{person}{Gustavo Alonso}.} \bibinfo{year}{2020}\natexlab{}.
\newblock \showarticletitle{{StRoM: Smart remote memory}}.
\newblock \bibinfo{journal}{\emph{Proceedings of the 15th European Conference
  on Computer Systems, EuroSys 2020}} (\bibinfo{year}{2020}).
\newblock
\showISBNx{9781450368827}
\urldef\tempurl%
\url{https://doi.org/10.1145/3342195.3387519}
\showDOI{\tempurl}


\bibitem[\protect\citeauthoryear{Tsai, Payer, and Zhang}{Tsai
  et~al\mbox{.}}{2019}]%
        {Pythia}
\bibfield{author}{\bibinfo{person}{Shin~Yeh Tsai}, \bibinfo{person}{Mathias
  Payer}, {and} \bibinfo{person}{Yiying Zhang}.}
  \bibinfo{year}{2019}\natexlab{}.
\newblock \bibinfo{booktitle}{\emph{{Pythia: Remote oracles for the masses}}}.
\newblock 693--710 pages.
\newblock
\showISBNx{9781939133069}
\urldef\tempurl%
\url{https://www.usenix.org/conference/usenixsecurity19/presentation/tsai}
\showURL{%
\tempurl}


\bibitem[\protect\citeauthoryear{Tsai and Zhang}{Tsai and Zhang}{2017}]%
        {literdma}
\bibfield{author}{\bibinfo{person}{Shin~Yeh Tsai} {and} \bibinfo{person}{Yiying
  Zhang}.} \bibinfo{year}{2017}\natexlab{}.
\newblock \showarticletitle{{LITE Kernel RDMA Support for Datacenter
  Applications}}.
\newblock \bibinfo{journal}{\emph{SOSP 2017 - Proceedings of the 26th ACM
  Symposium on Operating Systems Principles}} (\bibinfo{year}{2017}),
  \bibinfo{pages}{306--324}.
\newblock
\showISBNx{9781450350853}
\urldef\tempurl%
\url{https://doi.org/10.1145/3132747.3132762}
\showDOI{\tempurl}


\bibitem[\protect\citeauthoryear{Wang, Ma, Liu, Li, Ruan, Nguyen, Bond,
  Netravali, Kim, and Xu}{Wang et~al\mbox{.}}{2020}]%
        {semeru}
\bibfield{author}{\bibinfo{person}{Chenxi Wang}, \bibinfo{person}{Haoran Ma},
  \bibinfo{person}{Shi Liu}, \bibinfo{person}{Yuanqi Li},
  \bibinfo{person}{Zhenyuan Ruan}, \bibinfo{person}{Khanh Nguyen},
  \bibinfo{person}{Michael~D. Bond}, \bibinfo{person}{Ravi Netravali},
  \bibinfo{person}{Miryung Kim}, {and} \bibinfo{person}{Guoqing~Harry Xu}.}
  \bibinfo{year}{2020}\natexlab{}.
\newblock \bibinfo{booktitle}{\emph{{Semeru: A memory-disaggregated managed
  runtime}}}.
\newblock 261--280 pages.
\newblock
\showISBNx{9781939133199}
\urldef\tempurl%
\url{https://www.usenix.org/conference/osdi20/presentation/wang}
\showURL{%
\tempurl}


\bibitem[\protect\citeauthoryear{Zhu, Eran, Firestone, Guo, Lipshteyn, Liron,
  Padhye, Raindel, Yahia, and Zhang}{Zhu et~al\mbox{.}}{2015}]%
        {rocev2}
\bibfield{author}{\bibinfo{person}{Yibo Zhu}, \bibinfo{person}{Haggai Eran},
  \bibinfo{person}{Daniel Firestone}, \bibinfo{person}{Chuanxiong Guo},
  \bibinfo{person}{Marina Lipshteyn}, \bibinfo{person}{Yehonatan Liron},
  \bibinfo{person}{Jitendra Padhye}, \bibinfo{person}{Shachar Raindel},
  \bibinfo{person}{Mohamad~Haj Yahia}, {and} \bibinfo{person}{Ming Zhang}.}
  \bibinfo{year}{2015}\natexlab{}.
\newblock \showarticletitle{{Congestion Control for Large-Scale RDMA
  Deployments}}.
\newblock \bibinfo{journal}{\emph{Computer Communication Review}}
  \bibinfo{volume}{45}, \bibinfo{number}{4} (\bibinfo{year}{2015}),
  \bibinfo{pages}{523--536}.
\newblock
\showISBNx{9781450335423}
\showISSN{19435819}
\urldef\tempurl%
\url{https://doi.org/10.1145/2785956.2787484}
\showDOI{\tempurl}


\end{thebibliography}

\end{document}